\documentclass[aps,prl,10pt,twocolumn,showpacs,preprintnumbers,amsmath,amssymb,showkeys,citeautoscript]{revtex4-1}
\usepackage{pdftexcmds}
\usepackage[version=3]{mhchem}
\usepackage{siunitx}
\usepackage{xcolor}
\usepackage{graphicx}
\usepackage{dcolumn}
\usepackage{bm}
\usepackage{amsmath}
\usepackage{tikz}
\usepackage{pgfplots}
\usepackage{chemformula} 
\usepackage[T1]{fontenc} 

\begin{document}

\title{Dramatic enhancement of visible-light absorption in \ce{TiO2} by adding Bi}

\author{Fernando P. Sabino}
\affiliation{Department of Materials Science and Engineering, University of Delaware, Newark, Delaware 19716, USA}
\affiliation{Centro de Ci\^encias Naturais e Humanas, Universidade Federal do ABC, Santo Andr\'e, S\~ao Paulo, 09210-580, Brazil}
\author{Anderson Janotti}
\affiliation{Department of Materials Science and Engineering, University of Delaware, Newark, Delaware 19716, USA}

\begin{abstract}

\ce{TiO2} is a wide band-gap semiconductor that has been intensively investigated for photocatalysis and water-spiting.  
However, weak light absorption in the visible region of the spectrum poses stringent limitation to its practical application. Doping of \ce{TiO2} with \ce{N} or transition-metal impurities has been explored to shift the onset of optical absorption to the visible region, yet with limited success. Based on hybrid density functional calculations, we propose adding \ce{Bi} to \ce{TiO2}, in the form of dilute \ce{Ti_{1-x}Bi_{x}O2} alloys, to efficiently shift the optical absorption to the visible region. Compared to \ce{N}, \ce{Bi} introduces an intermediate valence band that is significantly higher in the band gap, and leaves the conduction band almost unchanged, leading to a remarkable redshift in the absorption coefficient to cover almost all the visible-light spectrum. Comparing formation enthalpies, our results show that adding \ce{Bi} costs significantly less energy than \ce{N} in oxidizing conditions, and that \ce{Ti_{1-x}Bi_{x}O2} might make a much more efficient photocatalyst than \ce{TiO_{2-y}N_{y}} for water splitting.

\end{abstract}


\maketitle

\section{Introduction} \label{sec:Introduction}

Solar energy is an abundant, clean, and renewable energy resource, and our ability to efficiently harness and use it is key for an energy sustainable future \cite{Braff2016}. Artificial photosynthesis, and photocatalysis in general,  is a promising technology to harvest and store solar energy, producing hydrogen or hydrocarbons via water splitting or \ce{CO2} transformation using a photoelectrochemical device \cite{Ye2018,Jafari2016}.  It remains a significant challenge to fabricate an efficient and stable solar-energy conversion device. Controlling the semiconductor properties in these devices is of primary concern in the development of new materials for solar-energy conversion. Among the semiconductor materials that have been explored as photoelectrodes, \ce{TiO2} stands out as being composed of earth-abundant and non toxic elements, and being photochemically stable under acidic and basic conditions \cite{Fujishima_37_1972,Ni_401_2007,Jiang2017,Cheng2021}. However, the large band gap of 3.0 eV for rutile and 3.2 eV for anatase severely limit the performance of \ce{TiO2} since only 5\% of the solar spectrum can in principle be  utilized\cite{Pascual_5606_1978,Tang_847_1993,Jiang2017}.


Attempts have been made to extend light absorption in \ce{TiO2} to the visible region by adding impurities (doping) or forming dilute alloys, while maintaining the photochemical stability and low cost. Adding \ce{N} has been considered one of the most effective way to bring light absorption in \ce{TiO2} to the visible range  \cite{Sato_126_1986,Asahi_269_2001,Chambers_27_2007,Cheung_1754_2007,Varley_2343_2011,Piskunov2015,Cheng2021}. The 2$p$ orbitals of the \ce{N} substituting for O couple with the O 2$p$ orbitals and lead to N-related bands above the original valence band, resulting in photoabsorption in the upper part of the visible spectrum. Similar effects have also been proposed to occur for \ce{C} and \ce{S} additions\cite{Piskunov2015,Umezawa2008}. 
Transition-metal and noble-metal additions, such as 
Cr, Co, V, Fe, Au, Ag, Cu, Pt and Pd \cite{Li_2381_2001,Fuerte_2718_2001,Dvorabiva_91_2002,Pelaez_331_2012,Cheng2021}, have also been proposed to bring the optical absorption in \ce{TiO2} into the visible region.  In the cases that have been tried in the laboratory \cite{Sato_126_1986,Asahi_269_2001,Chambers_27_2007,Cheung_1754_2007,Li_2381_2001,Pelaez_331_2012,Li_2381_2001,Pelaez_331_2012}, the reported enhancements in the visible-light absorption are not substantial, and limited to wavelengths longer than 450 nm (i.e., photon energies higher than 2.75 eV).
Such slight improvements were attributed in part to the low solubility and insufficient redshift of the band gap in the case of \ce{N} \cite{Sato_126_1986,Asahi_269_2001,Chambers_27_2007,Cheung_1754_2007}, or to introducing localized in-gap states close to the conduction band and incorporating on the surface and blocking the catalytic sites in the case of transition metals \cite{Li_2381_2001,Pelaez_331_2012}; 
 visible-light responsiveness of \ce{TiO2} with noble-metal additions was attributed to noble-metal related surface plasmons \cite{Pelaez_331_2012} instead of absorption in the \ce{TiO2}.

More recently, addition of post transition metals to \ce{TiO2}, such as \ce{Bi}, has also been considered, aiming at enhancing visible-light absorption and improving its photocatalytic efficiency
\cite{Wu2009,Sajjad_13795_2010,Kang_1371_2011,Ming_04CJ01_2017,Xu_530_2017}. Earlier work on 
the effects of adding Bi to \ce{TiO2} reported an increase of a factor of 10 in the hydrogen photo-generation rate and a remarkable photocurrent enhancement, with optimum results obtained for 1\% mol Bi content \cite{Wu2009}. However, it remains unclear if these improvements came from Bi at the surface or Bi incorporated in the bulk. Subsequent experiments of Bi-added \ce{TiO2} thin films and nanoparticles also reported improvements in photocatalytic efficiency, yet the proposed mechanisms either involved  the assumption of a Bi-related band near the \ce{TiO2} conduction band \cite{Sajjad_13795_2010,Kang_1371_2011} or Bi-metal/\ce{Bi2O3} formation at the surface \cite{Xu_530_2017}. The microscopic mechanism, local structure, and effects of Bi on the electronic structure of \ce{TiO2} are yet to be resolved. 
  
Inspired by these earlier promising results, we performed density functional theory (DFT) and hybrid functional calculations 
to study the structural, electronic, and optical properties of dilute \ce{Ti_{1-x}Bi_{x}O2} alloys, in rutile and anatase phases. For comparison, we also performed calculations for dilute \ce{TiO_{2-y}N_{y}} alloys, focusing on the formation enthalpy of these two systems as function of \ce{Bi} and \ce{N} concentrations, and their electronic and optical properties. We find that \ce{Bi} introduces a partially occupied intermediate valence band, detached from the  original O 2$p$ valence band, lying almost in the middle of the band gap, leaving the conduction band unchanged. This intermediate valence band leads to high absorption coefficients in the visible range, making \ce{Ti_{1-x}Bi_{x}O2} much more robust for visible-light water splitting than \ce{TiO_{2-y}N_{y}} alloys.


\section{Computational Approach}

The calculations are based on density functional theory \cite{Hohenberg_B864_1964,Kohn_A1133_1965} within the Perdew, Burke, and Ernzerhof 
exchange and correlation functional revised for solids (PBEsol) \cite{Perdew_136406_2008} and the
 hybrid functional of Heyd, Scuseria, and Ernzerhof (HSE06) \cite{Heyd_7274_2004,Heyd_219906_2006},
 implemented with projected augmented wave (PAW) potentials \cite{Blochl_17953_1994} in the VASP code \cite{Kresse_13115_1993,Kresse_11169_1996}. 
The stress tensor and the atomic forces were relaxed using PBEsol with 
a cutoff energy of \SI{620}{\electronvolt} for the plane-wave basis set. We employed 
a \textbf{k}-point mesh of $5{\times}5{\times}9$ for integrations over the Brillouin 
zone of the 6-atom primitive cell of rutile \ce{TiO2}, and maintain the same \textbf{k}-mesh density for anatase and the supercells containing \ce{Bi} or \ce{N}.

Since PBEsol severely underestimates band gaps, we employ the 
HSE06 hybrid functional to describe the electronic properties of rutile and anatase \ce{TiO2}, \ce{Ti_{1-x}Bi_{x}O2} and 
\ce{TiO_{2-y}N_{y}}. Under this approximation, the exchange potential 
is divided in short and long-range parts by a screening parameter $\omega = 0.206$\AA$^{-1}$ \cite{Heyd_7274_2004,Heyd_219906_2006}. 
In the short-range part, non-local Hartree-Fock exchange is mixed with semi-local PBE exchange \cite{Perdew_3865_1996} in a ratio of 25\%/75\%; the long-range part is described by the PBE functional. For the electronic
structure calculations, with the HSE06 hybrid functional, the cutoff energy was reduced to \SI{470}{\electronvolt}.  The density of states (DOS) for the primitive cells and the supercells representing the alloys were calculated using $\Gamma$-centered \textbf{k}-point meshes that are equivalent to a $11{\times}11{\times}17$ mesh for the rutile \ce{TiO2} 6-atom primitive cell.
 

The dilute alloys in the rutile and anatase phases, $r$-\ce{Ti_{1-x}Bi_{x}O2}, $a$-\ce{Ti_{1-x}Bi_{x}O2}, $r$-\ce{TiO_{2-y}N_{y}}, and $a$-\ce{TiO_{2-y}N_{y}}, were simulated using special quasi-random structures \cite{Zunger1990,VandeWalle2013} to represent random arrangements of \ce{Bi} or \ce{N}. For the rutile phase we use a supercell of 192 atoms where we replaced 1, 2, and 3 \ce{Ti} with \ce{Bi} atoms to simulate concentrations of 1.6\%, 3.1\%, and 4.7\%; equivalently we replaced 2, 4, and 6 \ce{O} with \ce{N} atoms. For the anatase phase, we used a supercell of 108 atoms where we replaced 1 and 2 \ce{Ti} with \ce{Bi} atoms to simulate concentrations of 2.8\% and 5.6\%; equivalently we replaced 2 and 4 \ce{O} with \ce{N} atoms.

Since \ce{Bi} and \ce{N} are aliovalent species in \ce{TiO2}, the calculations for the density of states and optical properties of \ce{Ti_{1-x}Bi_{x}O2} and \ce{TiO_{2-y}N_{y}} alloys were performed for charge compensated closed shell systems, i.e., with the intermediate valence bands completely filled, assuming that, in practice, donor defects such as oxygen vacancies will be present as compensation centers.
This avoids the difficulties of having to calculate dielectric matrices for metallic systems.
The optical properties were computed with the tetrahedral smearing method, phonon assisted transitions and exciton effects were neglected, as these effects will not affect our conclusions.

Finally, due to the high computation cost to calculate the dielectric matrix with required high-density \textbf{k}-point mesh for the alloy supercells using HSE06, we performed the calculations with PBEsol and 
used a scissors operator to the PBEsol results and shifted the optical absorption coefficient based on the difference between the band gap obtained in HSE06 and PBEsol. This approach is expected to not affect the intensity of the absorption coefficient near the band gap or elsewhere.

\section{Results and Discussion}

\subsection{Structural properties of \ce{Ti_{1-x}Bi_{x}O2} and \ce{TiO_{2-y}N_{y}} alloys and their formability}

\begin{figure}
\centering
\includegraphics[width=1.00\linewidth]{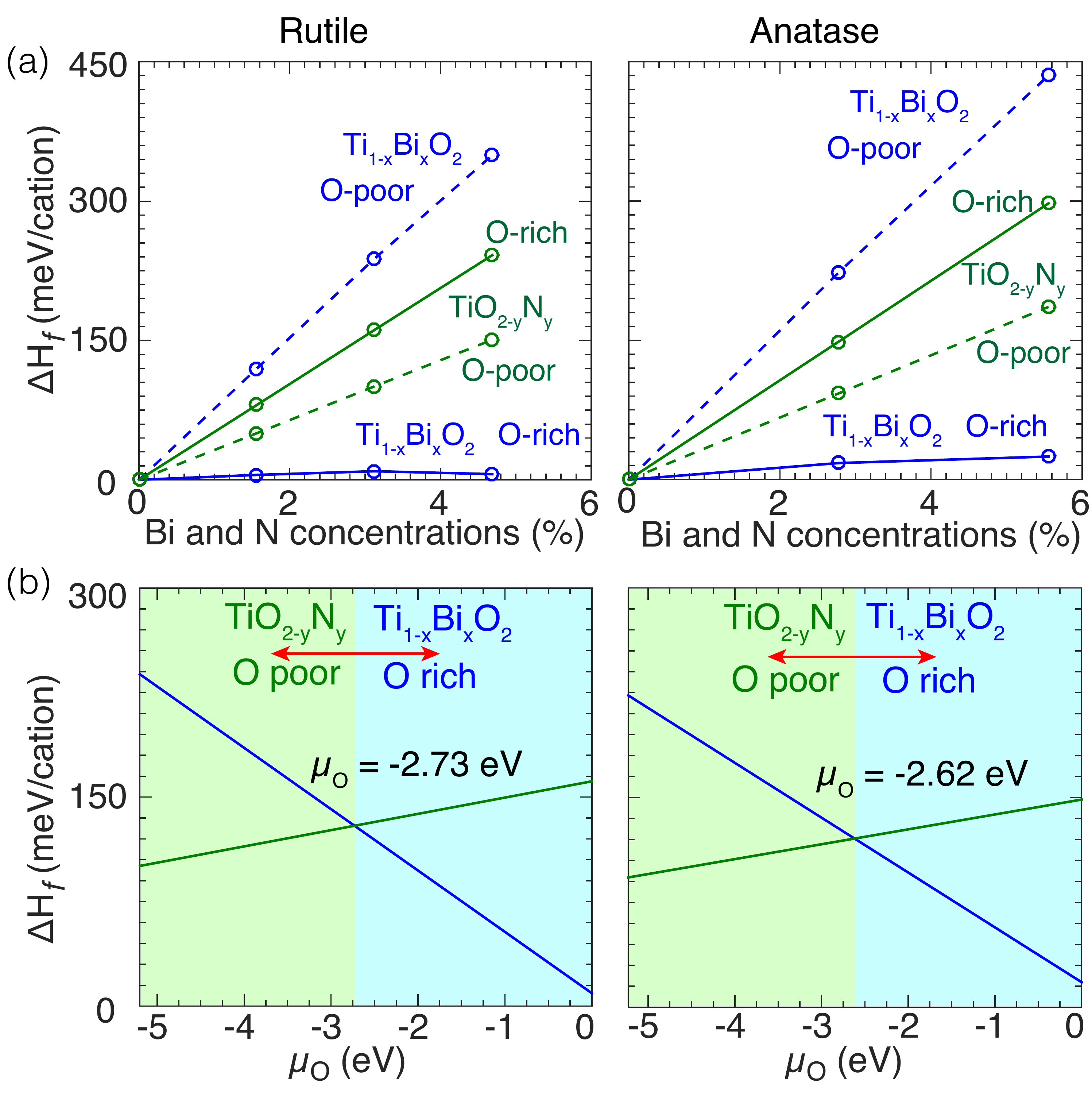}
\caption{Formation enthalpy ($\Delta H_f$) of  \ce{Ti_{1-x}Bi_{x}O2} and \ce{TiO_{2-y}N_{y}} alloys.
(a) Formation enthalpy as function of \ce{Bi} and \ce{N} content for the alloys in rutile ($r$-) and anatase ($a$-) phases.
The continuous lines refer to O-rich limit condition, while the 
dashed lines refer to the O-poor limit condition. (b) Formation enthalpy as function of  
chemical potential $\mu_{\rm O}$ for \ce{Bi} and \ce{N} concentrations of $3.13\%$ for $r$-\ce{Ti_{1-x}Bi_{x}O2} and $r$-\ce{TiO_{2-y}N_{y}}, and $2.8\%$ for the alloys in the anatase phase.
The crossing point in $\mu_{\rm O}$ indicates the chemical potential value above which \ce{Ti_{1-x}Bi_{x}O2} alloys have lower formation enthalpy than \ce{TiO_{2-y}N_{y}}, showing that the former are more more favorable to form in more oxidizing conditions.
}
\label{Fig_1}
\end{figure}

The most common phase of bulk \ce{TiO2} is rutile ($r$-\ce{TiO2}), whereas anatase ($a$-\ce{TiO2}) is mostly found in thin-film and nanostructure forms.
$r$-\ce{TiO2} belongs to $P4_2/mnm$ space group and contains 6 atoms in the primitive cell,
while $a$-\ce{TiO2} belongs to $I4_1/amd$ space group and contains 12 atoms in the primitive cell.
In both phases, each \ce{Ti} atom is surrounded by six \ce{O} atoms forming edge-sharing octahedra, which are almost perfect in rutile and highly distorted in anatase.
Each \ce{O} atom is bonded to three \ce{Ti} atoms in planar configurations. 
Based on atomic radii \cite{Slater1964}, \ce{Bi} (1.6 \AA)
is expected to incorporate on the \ce{Ti} (1.4 \AA) sites, while \ce{N} (0.65 \AA) is expected to substitute for
\ce{O} (0.60 \AA).  

\begin{table}[t!]
\centering
\caption{Lattice parameters $a_0$ and $c_0$ and volume of $r$-\ce{TiO2} (rutile), $a$-\ce{TiO2} (anatase) and
\ce{Ti_{1-x}Bi_{x}O2} and \ce{TiO_{2-y}N_{y}} alloys for different \ce{Bi} and \ce{N} concentrations ($x$ or $y$).
The experimental data for $r$-\ce{TiO2} and $a$-\ce{TiO2} from Refs.~\cite{Wyckoff_1963,Horn_273_1972} are also listed. }
\label{lattice}
\begin{tabular}{lcccc}
Material                          &~$x$ or $y$       & $a_0$        & $c_0$       & Volume         \\
                                  &($\%$)            & (\AA)      & (\AA)     & (\AA$^3$/f.u.)  \\
\hline
$r$-\ce{TiO2}                     &                  & 4.584         & 2.937        & 30.86     \\
$r$-\ce{TiO2} (exp.)              &                  & 4.594         & 2.959        & 31.22     \\
$r$-\ce{Ti_{1-x}Bi_{x}O2}         & ~$1.6$          & 4.593         & 2.945        & 31.06     \\
$r$-\ce{Ti_{1-x}Bi_{x}O2}         & ~$3.1$          & 4.604         & 2.952        & 31.28     \\
$r$-\ce{Ti_{1-x}Bi_{x}O2}         & ~$4.7$          & 4.617         & 2.960        & 31.55     \\
$r$-\ce{TiO_{2-y}N_{y}}           & ~$1.6$          & 4.584         & 2.939        & 30.87     \\
$r$-\ce{TiO_{2-y}N_{y}}           & ~$3.1$          & 4.585         & 2.939        & 30.90     \\
$r$-\ce{TiO_{2-y}N_{y}}           & ~$4.7$          & 4.587         & 2.940        & 30.93     \\
\hline
$a$-\ce{TiO2}                     &                  & 3.765         & 9.538        & 33.81     \\
$a$-\ce{TiO2} (exp.)              &                  & 3.784         & 9.515        & 34.06     \\
$a$-\ce{Ti_{1-x}Bi_{x}O2}         & ~$2.8$          & 3.782         & 9.579        & 34.25     \\
$a$-\ce{Ti_{1-x}Bi_{x}O2}         & ~$5.6$          & 3.800         & 9.624        & 36.73     \\
$a$-\ce{TiO_{2-y}N_{y}}           & ~$2.8$          & 3.767         & 9.543        & 33.86     \\
$a$-\ce{TiO_{2-y}N_{y}}           & ~$5.6$          & 3.768         & 9.545        & 33.88     \\

\end{tabular}
\end{table}

The calculated lattice parameters for $r$-\ce{TiO2} and $a$-\ce{TiO2}, listed in Table~\ref{lattice},
are in good agreement with the experimental data \cite{Wyckoff_1963,Horn_273_1972}.
Adding \ce{Bi} to \ce{TiO2}, with each \ce{Bi} substituting on a \ce{Ti} site, leads to 
a sizable increase in lattice parameters, attributed to the large atomic radius of Bi compared to that of Ti.
In contrast, adding \ce{N}, replacing \ce{O}, leads to only slight increase in lattice parameters and in the volume per formula unit.
For example, for \ce{N} concentration of $y=5.6\%$ in $a$-\ce{TiO_{2-y}N_{y}}, we find a volume expansion of only $0.20\%$,
while for \ce{Bi} concentration of $x=5.6\%$ in $a$-\ce{Ti_{1-x}Bi_{x}O2}, the volume increases by $8.6\%$.

\begin{figure*}
\centering
\includegraphics[width=1.00\linewidth]{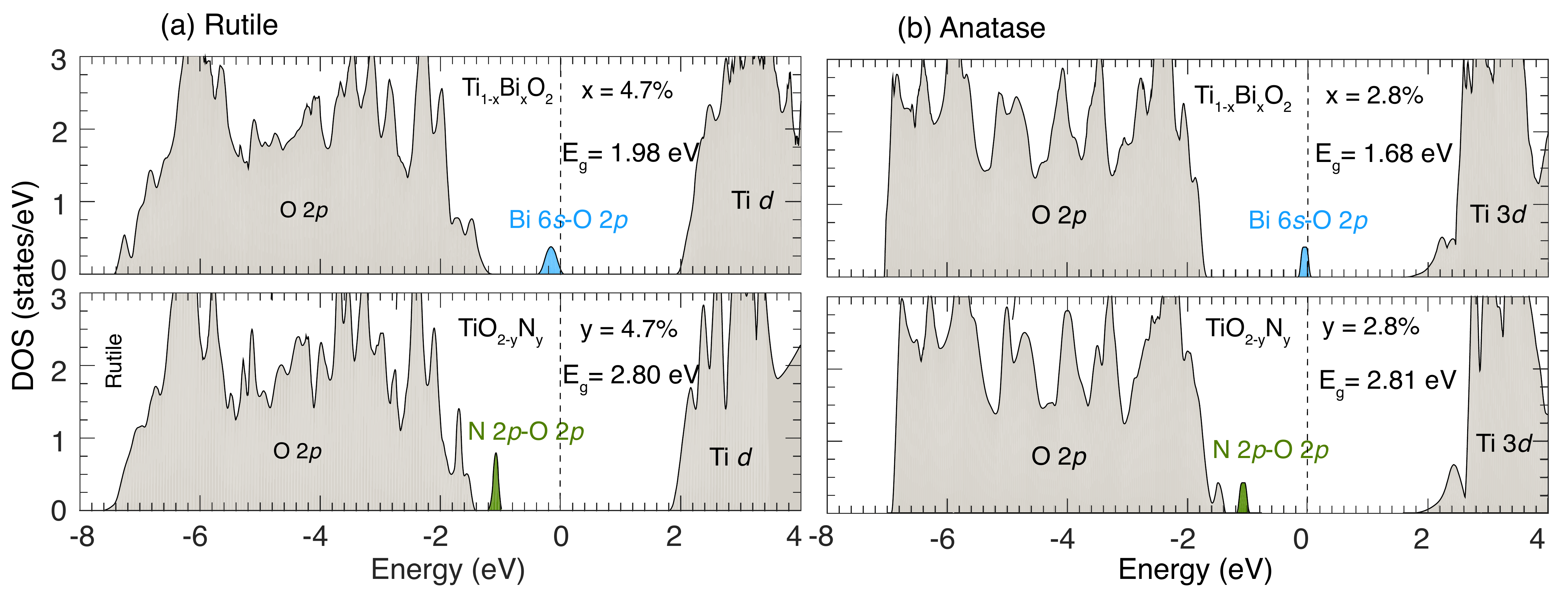}
\caption{
Electronic density of states (DOS) for dilute \ce{Ti_{1-x}Bi_{x}O2} and 
\ce{TiO_{2-y}N_{y}} alloys in (a) rutile and (b) anatase phases.  The zero in energy axes are set to the 
top of the intermediate valence band in the \ce{Ti_{1-x}Bi_{x}O2} alloys. The DOS of 
\ce{Ti_{1-x}Bi_{x}O2} and \ce{TiO_{2-y}N_{y}} in the same phase were aligned using the low
lying O 2$s$ bands located around $-19$ eV. The results show that \ce{Ti_{1-x}Bi_{x}O2} alloys 
show smaller energy differences between the intermediate valence band and the conduction band than  
\ce{TiO_{2-y}N_{y}} alloys.  In these calculations electrons were added to have fully occupied intermediate 
valence bands with \ce{Bi} and \ce{N}, representing electronic compensated dilute alloys without explicitly adding the compensating centers, such as oxygen vacancies\cite{Janotti_085212_2010} for example. 
}
\label{Fig_2}
\end{figure*}

Neglecting the charge state of the Bi addition, the formation enthalpy of the dilute \ce{Ti_{1-x}Bi_{x}O2} alloys  
are calculated using:
\begin{align}
\Delta H_f ({\rm Ti}_{1-x}{\rm Bi}_{x}{\rm O}_2) & = E_{tot}({\rm Ti}_{1-x}{\rm Bi}_{x}{\rm O}_2) - E_{tot}({\rm TiO}_2)  \nonumber \\
                                               & +n [E_{tot}({\rm Ti}) - E_{tot}({\rm Bi})] \nonumber \\
                                               & + n (\mu_{\rm Ti} - \mu_{\rm Bi}),
\end{align}
where $E_{tot}({\rm Ti}_{1-x}{\rm Bi}_{x}{\rm O}_2)$ is the total energy of the supercell representing the alloy, $E_{tot}({\rm TiO}_2)$
is the total energy of pristine \ce{TiO2} using the same supercell size.  The chemical potentials $\mu_{\rm Ti}$ and $\mu_{\rm Bi}$ are referenced to the total energies of \ce{Ti} and \ce{Bi} bulk metallic phases ($E_{tot}({\rm Ti})$ and $E_{tot}({\rm Bi})$), with $\mu_{\rm Ti}\leq 0$ and $\mu_{\rm Bi} \leq 0$.
Similar expression is used to calculate the formation enthalpy of the dilute \ce{TiO_{2-y}N_{y}} alloys. The variation of the chemical potentials $\mu_{\rm Ti}$, $\mu_{\rm O}$, $\mu_{\rm Bi}$, and $\mu_{\rm N}$ are restricted by the stability of \ce{TiO2} and the formation of the secondary phases \ce{Bi2O3} and \ce{TiN}.
For example, in the O-rich limit condition we have: $\mu_{\rm O}=0$, $\mu_{\rm Ti}=\Delta H_f({\rm TiO}_2)$, 
$\mu_{\rm Bi}=\frac{1}{2}\Delta H_f({\rm Bi}_2{\rm O}_3)$, and $\mu_{\rm N}=0$; in the O-poor limit we have:
$\mu_{\rm O}=\frac{1}{2}\Delta H_f({\rm TiO}_2)$, $\mu_{\rm Ti}=0$, $\mu_{\rm Bi}=0$, and $\mu_{\rm N}=\Delta H_f({\rm TiN})$.
$\Delta H_f({\rm TiO}_2)$, $\Delta H_f({\rm Bi}_2{\rm O}_3)$, and $\Delta H_f({\rm TiN})$ are the formation enthalpy of
\ce{TiO2}, \ce{Bi2O3}, and \ce{TiN}, respectively.
The results for alloy formation enthalpy ($\Delta H_f$) as function of concentration of \ce{Bi} and \ce{N} in the O-rich and O-poor limit conditions are shown in Fig.~\ref{Fig_1}(a).

The chemical potential $\mu_{\rm O}$ plays important role in the incorporation of \ce{Bi}
and \ce{N} in \ce{TiO2}. Since \ce{Bi} occupies the \ce{Ti} site, its incorporation is most favorable in O-rich (Ti-poor) conditions, while the incorporation of \ce{N}  is most favorable in O-poor (Ti-rich) conditions since it sits on the \ce{O} site. 
As seen in Fig.~\ref{Fig_1}(a), we find that the formation enthalpy of the dilute \ce{Ti_{1-x}Bi_{x}O2} alloys varies over a wider range from O-rich to O-poor limit conditions compared to the \ce{TiO_{2-y}N_{y}} alloys. 
A crossing in the formation enthalpy plot as function of $\mu_{\rm O}$ is thus expected. This crossing indicates the $\mu_{\rm O}$ value above which the formation enthalpy of the \ce{Ti_{1-x}Bi_{x}O2} alloy is  lower than that of the \ce{TiO_{2-y}N_{y}} for the same \ce{Bi} and \ce{N} concentration.  Since the formation enthalpy varies linearly with concentration in the dilute regime considered here, the $\mu_{\rm O}$ value at the crossing does not depend on the \ce{Bi} and \ce{N} content, but it is different for the two phases.  We find that $\mu_{\rm O}=-2.73$ eV for rutile and $-2.62$ eV for anatase. 
Considering typical growth conditions of \ce{TiO2} thin films by molecular beam epitaxy \cite{Chambers_27_2007,Cheung_1754_2007}, for example, with temperature in the range of 400$-$700$^{\circ}$C and pressure of $10^{-3}-10^{-7}$ torr, $\mu_{\rm O}$ for \ce{O2} gas falls in the range of $-1.7$ eV to $-0.9$ eV, being closer to the O-rich limit, and favoring Bi incorporation. 
The results in Fig.~\ref{Fig_1}(a) also show that it cost less energy to incorporate \ce{Bi} on the octahedral chemical environment of rutile than on the distorted octahedral environment of the anatase phase. In contrast, the formation enthalpy for the incorporation of \ce{N} in $r$-\ce{TiO2} and $a$-\ce{TiO2} are almost the same for a given value of $\mu_{\rm O}$, which we attribute to the similarity between the anion chemical environment  in rutile and anatase.

Both \ce{Ti_{1-x}Bi_{x}O2} and \ce{TiO_{2-y}N_{y}} alloys have been demonstrated
\cite{Sato_126_1986,Asahi_269_2001,Chambers_27_2007,Cheung_1754_2007,Sajjad_13795_2010,Kang_1371_2011,Ming_04CJ01_2017,Xu_530_2017},
with \ce{Bi} and \ce{N}  concentrations of up to a few atomic percent. 
For \ce{Bi}, it was found a maximum solubility around $5\%$ before formation of a secondary phase occurs\cite{Ming_04CJ01_2017}.
These dilute concentrations are consistent with the values of formation enthalpy shown in Fig.~\ref{Fig_1}.  Our results show a large variation of the formation enthalpy of \ce{Ti_{1-x}Bi_{x}O2} with $\mu_{\rm O}$, and indicate that for
maximizing the \ce{Bi} concentration, O-rich growth or deposition conditions should be employed.

\subsection{Electronic structure and optical properties of \ce{Ti_{1-x}Bi_{x}O2} and \ce{TiO_{2-y}N_{y}} alloys}

For the effects of adding \ce{Bi} on the electronic and optical properties of \ce{TiO2}, our results show that \ce{Bi} leads to remarkably larger
redshit in band gap and optical absorption than \ce{N}.  The calculated density of states (DOS) of dilute  \ce{Ti_{1-x}Bi_{x}O2} and \ce{TiO_{2-y}N_{y}} with ${\rm x=y=4.7}\%$ for rutile and ${\rm x=y=2.8}\%$ for anatase are shown in Fig~\ref{Fig_2}.

\begin{figure}
\centering
\includegraphics[width=1.00\linewidth]{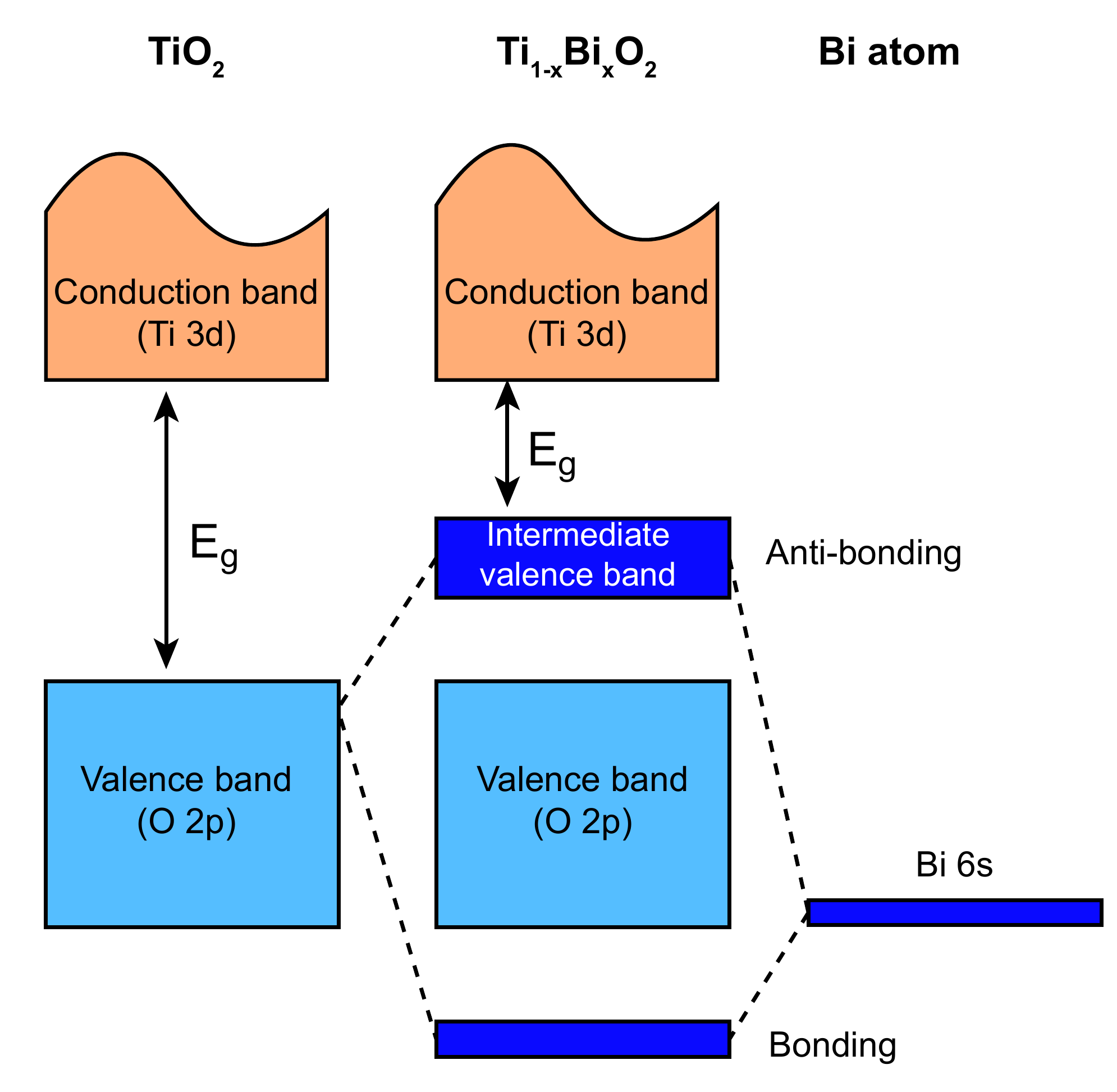}
\caption{
Schematic representation of the coupling between the \ce{Bi} 6$s$ and \ce{O} 2$p$ forming the intermediate valence band in
dilute \ce{Ti_{1-x}Bi_{x}O2} alloys, considerably reducing the excitation energy to the conduction band, yet leaving the position of the conduction band unchanged.}
\label{Fig_3}
\end{figure}

In \ce{TiO2}, the band gap of $\sim$3 eV separates the occupied valence band, derived mostly from \ce{O} 2$p$ orbitals, from the unoccupied conduction band derived mostly from the \ce{Ti} 3$d$ orbitals.
For $r$-\ce{TiO2}, the calculated band gap is $3.10$ eV, and 3.35 eV for $a$-\ce{TiO2}, compared to the
 experimental values of $3.02$ eV \cite{Pascual_5606_1978} and $3.20$ eV, respectively.
When \ce{Bi} is incorporated into \ce{TiO2}, forming a dilute \ce{Ti_{1-x}Bi_{x}O2} alloy, a partially occupied intermediate band is created and located in the band gap, closer to the valence band. This intermediate band is derived from 
the coupling between the \ce{O} 2$p$ and the low lying \ce{Bi} 6$s$ bands--- the later is located between $-9$ and $-10$ eV below the top of the valence band, as schematically shown in Fig.~\ref{Fig_3}. The partially occupied intermediate valence band in \ce{TiO_{2-y}N_{y}} alloys, for comparison, is much closer to the valence band, visibly modifying the top of the \ce{O} 2$p$ band,
where a small shoulder in the DOS can be distinguished.

The intermediate valence band significantly reduces the excitation energy to the conduction band as
indicated in Fig.~\ref{Fig_2} and  Fig.~\ref{Fig_3}, and this effect is remarkably stronger in \ce{Ti_{1-x}Bi_{x}O2} than in \ce{TiO_{2-y}N_{y}}.  Note than the distance between the  \ce{O} 2$p$ and the \ce{Ti} 3$d$ bands remains almost unchanged, indicating that adding \ce{Bi} or \ce{N} only affects the highest occupied bands and does not change the position of the conduction band with respect to the vacuum level. This is a desirable effect in photoelectrochemical devices based on \ce{TiO2}, enabling visible-light absorption while keeping the conditions that favor the redox potentials for water splitting \cite{Jiang2017}.

\begin{figure}
\centering
\includegraphics[width=1.00\linewidth]{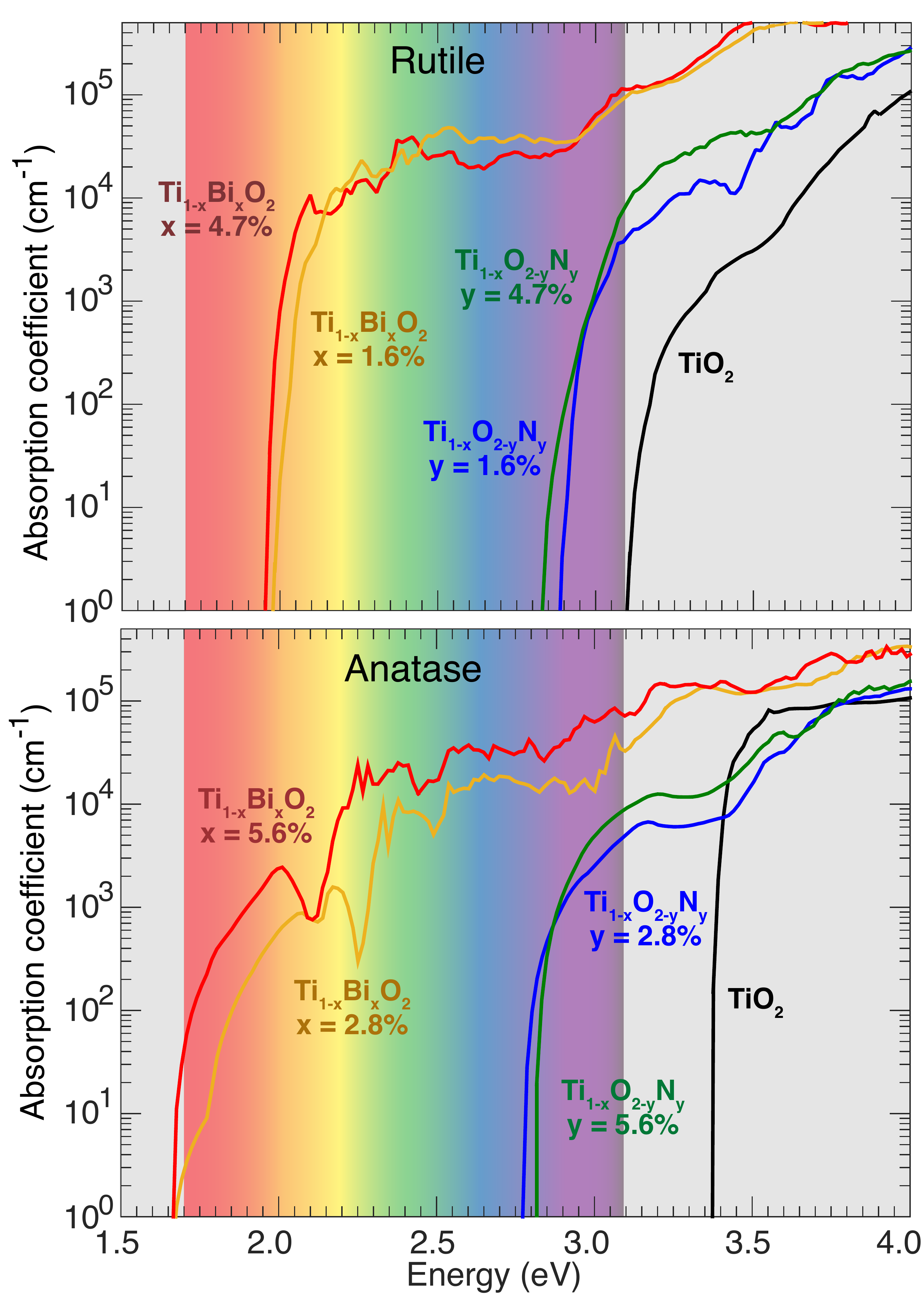}
\caption{
Calculated absorption coefficient (averaged over the three cartesian directions) as a function of photon 
energy for \ce{TiO2}, \ce{Ti_{1-x}Bi_{x}O2} and \ce{TiO_{2-y}N_{y}} with 
Bi and N concentrations of 1.6 and 4.7$\%$ in rutile and 2.8 and 5.6$\%$ in anatase crystal structure. The colored region indicate the visible-light spectrum.
}
\label{Fig_4}
\end{figure}

Our results also show that adding \ce{Bi} not only enables visible-light absorption 
over a wider range in the spectrum, but also leads to a significantly increase in
the absorption coefficient near the effective band gap (between the intermediate valence band and the conduction band),
compared to adding \ce{N}. The calculated absorption coefficients for \ce{Ti_{1-x}Bi_{x}O2}, \ce{TiO_{2-y}N_{y}}, and 
\ce{TiO2} for comparison, are shown in Fig.~\ref{Fig_4}. Having the system electronically compensated
facilitates these calculations, and the results were averaged over the three orthogonal directions,
which can be argued to better represent the absorption in polycrystalline films or nanostructures.

For bulk \ce{TiO2}, the optical band gap does not coincide with the fundamental band gap, 
as observed for several other oxides, including \ce{SnO2}, \ce{In2O3}, \ce{PbO2} 
\cite{Walsh_167402_2008,Scanlon_246402_2011,Sabino_205308_2015,Sabino_085501_2017}.
For $r$-\ce{TiO2}, the dipole transition from valence-band maximum to conduction-band 
minimum at $\Gamma$ is forbidden due to symmetry considerations; however,  transitions in the vicinity 
of the $\Gamma$ point are allowed (yet with amplitude smaller 
than $\sim$$10^{-4}$ cm$^{-1}$) and, due to the relatively small dispersion of the valence and conduction bands,
the onset in the optical absorption coefficient is slightly shifted, by $\sim$0.1 eV, to higher energies with respect to the fundamental band gap, as shown in Fig.~\ref{Fig_4}(a). 
For $a$-\ce{TiO2}, the dipole matrix element for the minimum-energy transition from valence to conduction band is also forbidden by symmetry, and due to larger dispersion of the valence and conduction bands, the optical gap is shifted to higher energies by $\sim$0.4 eV with respect to the fundamental band gap, as shown in Fig.~\ref{Fig_4}(b). These results corroborate the fact that \ce{TiO2} by itself is so inefficient for visible-light photoelectochemical processes.

For dilute \ce{TiO_{2-y}N_{y}} alloys, the \ce{N}-related intermediate valence band leads to a reduction in the effective band gap, 
indicating that the alloy absorbs visible light, yet restricted to photons of relatively high energies.
 As shown in Fig.~\ref{Fig_4}, $r$-\ce{TiO_{2-y}N_{y}} alloys start absorbing visible light
 at $\sim$2.7 eV (459 nm), while $a$-\ce{TiO_{2-y}N_{y}} alloys start at $\sim$2.6 eV (477 nm).
The position of the onset of optical absorption varies only slightly with \ce{N} concentration; 
nevertheless, the amplitude of the absorption coefficients 
increases with \ce{N} content as seen in Fig.~\ref{Fig_4}.
Experimentally, it is known that adding \ce{N} to $a$-\ce{TiO2} leads to visible light absorption
starting at around \SI{2.5} {\electronvolt} \cite{Sato_126_1986,Asahi_269_2001}.  This is in good agreement with our calculations considering that the calculated band gap of $a$-\ce{TiO2} using HSE06 is 0.15 eV higher than the experimental value.

In the case of dilute \ce{Ti_{1-x}Bi_{x}O2} alloys, the redshift in the absorption coefficient is significantly larger 
than in \ce{TiO_{2-y}N_{y}}. The \ce{Bi}-related intermediate valence band lies almost in the 
middle of \ce{TiO2} band gap, and the predicted onset of optical absorption occurs at 2.0 eV (620 nm)
in $r$-\ce{Ti_{1-x}Bi_{x}O2} and 1.7 eV (719 nm) in $a$-\ce{Ti_{1-x}Bi_{x}O2}.  It the later, it covers almost all the visible spectrum.
Similar to the \ce{N} case, the position of the onset of optical absorption does not change with \ce{Bi} content in the dilute regime considered here,
yet high amplitudes in the absorption coefficient near the threshold are obtained with higher \ce{Bi} concentrations.
Our results explain experimental observations of \ce{Bi}-doped \ce{TiO2} indicating a band gap of 2.05 eV (600 nm) \cite{Kang_1371_2011}, yet the microscopic mechanism has not been addressed.

The combination of high optical absorption in the visible for both rutile and anatase phases of
dilute \ce{Ti_{1-x}Bi_{x}O2} alloys, and the fact that adding \ce{Bi} does not affect the position of the conduction band
make these alloys promising candidates for photocatalysis and water splitting. The comparison with \ce{TiO_{2-y}N_{y}}
clearly shows the superiority of adding \ce{Bi} instead of \ce{N}, leading to significantly higher absorption and wider range in the visible,
almost reaching the IR region of the spectrum [Fig.~\ref{Fig_4}(b)].

\section{Conclusions}

In summary, using DFT and hybrid functional calculations we show that adding \ce{Bi} leads to significantly more efficient visible-light absorption than adding N to \ce{TiO2}. Our results for
$a$-\ce{Ti_{1-x}Bi_{x}O2} alloy with 5.6\% shown onset of optical absorption at $\sim$1.7 eV, which is near the upper limit of the IR spectrum, thus efficiently covering almost the whole visible region.  As with \ce{N}, adding \ce{Bi} does not affect the position of the conduction band, offering an optimum straddling of the redox potentials for water splitting. The incorporation of \ce{Bi} is predicted to be most favorable in oxidizing conditions in contrast to \ce{N} incorporation. Our results not only explain the available experimental data on \ce{Bi}-doped \ce{TiO2}, but also provide the microscopic mechanisms for the observed enhancement of visible-light absorption,
calling for further experimental efforts to study the stability and performance of \ce{Bi}-added \ce{TiO2} in photocatalysis and water splitting.

\section{Acknowledgments}

This work was supported by the NSF Early Career Award grant no.~DMR-1652994, the Extreme Science and Engineering Discovery Environment (XSEDE) supported by National Science Foundation grant number ACI-1053575, and the Information Technologies (IT) resources at the University of Delaware.  FPS acknowledges support from FAPESP grant no.~2019/21656-8. 


%

\end{document}